# Topological phase transition and tunable surface states in YBi

Ramesh Kumar[1], Mukhtiyar Singh[1*]

[1]Department of Applied Physics, Delhi Technological University, New Delhi-110042, India

[*]mukhtiyarsingh@dtu.ac.in; msphysik09@gmail.com

A unique co-existence of extremely large magnetoresistance (XMR) and topological characteristics in non-magnetic rare-earth monopnictides stimulating intensive research on these materials. Yttrium monobismuthide (YBi) has been reported to exhibit XMR up to $10^5$% but its Topological properties still need clarification. Here we use the hybrid density functional theory to probe the structural, electronic and topological properties of YBi in detail. We observe that YBi is topologically trivial semimetal at ambient pressure which is in accordance with reported experimental results. The topological phase transitions i.e., trivial to non-trivial are obtained with volumetric pressure of 6.5 GPa and 3% of epitaxial strain. This topological phase transitions are well within the structural phase transition of YBi (24.5 GPa). The topological non-trivial state is characterized by band inversions among *Y-d* band and *Bi-p* band near *Γ*- and *X-point* in the Brillouin zone. This is further verified with the help of surface band structure along (001) plane. The $Z_2$ topological invariants are calculated with the help of product of parities and evolution of Wannier charge centers. The occurrence of non-trivial phase in YBi with a relatively small epitaxial strain, which a thin film geometry can naturally has, might make it an ideal candidate to probe inter-relationship between XMR and non-trivial topology.

## I. INTRODUCTION

Topological Insulators (TIs) [1–3] with unique metallic surface states and insulating bulk band gap have grabbed high attention in condensed matter physics. These topological surface states are protected by time reversal symmetry (TRS) which separates them from conventional insulators [4–6]. These surface states are spin-momentum locked and robust against local perturbations [4,7]. Some semimetals are also reported to exhibit non-trivial topological states and can be divided into various categories e.g., Weyl, Dirac and nodal-line semimetals, etc. [8–10]. Breaking of TRS or spatial symmetry transforms Dirac semimetals into Weyl semimetals and a Dirac point splits into a pair of Weyl points [11]. The surface states in Dirac and Weyl semimetals can be described by Fermi arc [12–14], unlike the Dirac cone in case of Tis, which is due to the overlaps between the surface and the bulk states. These topological semimetals can be characterized by a band inversion in the bulk band structure and the $Z_2$ topological invariants which can be calculated with the help of product of parities at time reversal invariant momenta (TRIM) points as well as using Wilson loop [2,15,16]. Several rare-earth pnictides such as LnPn (Ln: rare earth element; Pn: As, Sb, Bi) have reported to exhibit topological states at ambient conditions [17–19]. Spin-orbit coupling (SOC) is a viable tool to tune a trivial topological material to non-trivial one. The SOC strength can be enhanced using pressure, strain, chemical doping, and alloying [19–22] etc. Amongst these, external pressure and strain are most suitable owing to their non-disruptive nature. The effect of volumetric pressure or epitaxial strain on any material reduces its bond length in the respective directions, bandwidth, and energy differences across the bands without affecting charge neutrality or stoichiometry. Various semimetals such as LaAs [20], LaSb [21], TmSb [23], TaAs [24] and YbAs [25] have been shown to become topologically non-trivial with pressure. Topological phase in LaSb [26], and SnTe [27] has also been observed under epitaxial strain and the same has been confirmed experimentally by angle-resolved photoemission spectroscopy (ARPES) in SnTe [27].

The YBi has been reported to be a perfectly compensated semimetal with XMR up to $10^5$% with equal hole and electron carrier concentration [28,29]. *First-principles* calculations within Perdew-Burke-Ernzerhof (PBE) functional have predicted YBi as a topologically non-trivial semimetal with band inversion near $\Gamma$-point [30]. Recent study using ARPES and *first-principles* calculations with mBJ functional has indicated that YBi is a topologically trivial semimetal [29]. This disparity raises a debate on true topological nature of YBi and deserves a thorough analysis as it may be useful to settle down the debate concerning whether XMR is caused by nontrivial topological aspects or complete electron- hole compensation.

This motivates us to systematically explore structural, electronic, and topological properties of YBi using density functional theory (DFT) with relatively accurate hybrid functional Heyd, Scuseria and Ernzerhof (HSE06). This functional had predicted accurate electronic states of other similar rare earth monopnctides and gave carrier densities in good agreements with experimental results [20,21,23,25,26]. We study the topological properties under external volumetric pressure and epitaxial strain and analysed the quantum phase transitions in detail. The topological states are observed with the help of band inversion in bulk band structure and surface Dirac cone projected on (001) plane. The $Z_2$ topological invariants are calculated using parities of wavefunctions TRIM points and evolution of Wannier charge centers (WCCs).

## II. COMPUTATIONAL DETAILS

All structural and electronic calculations of YBi with applied volumetric pressure and epitaxial strain was carried out in the framework of DFT [31–33] based *first-principles* approach with projector augmented wave (PAW) [33] technique as implemented in VASP code [34]. The PBE [35] functional followed by screened hybrid functional HSE06 [36,37] was used for more accurate results. The long range and short-range parts of HSE06 was employed with screening parameters as $\omega=0.201$ Å$^{-1}$. The PBE functional used for long range part but a mixing of 25% Fock exchange was carried out in short range part of HSE06 functional. The PAW potentials used for Y and Bi were having eleven valance electrons (i.e., $4s^24p^55s^14d^2$) and fifteen valance electrons (i.e., $5d^{10}6s^26p^3$), respectively. An optimized Monkhorst-Pack type k-mesh of 7×7×7 and kinetic energy cutoff of 340 eV were used to calculate the plane wave basis set. Gaussian smearing method was set at width of 0.001 eV for Fermi level broadening and all atomic positions were fully relaxed.

To apply the volumetric pressure and epitaxial stain, we were used eight atom cubic unit cell. Out of which, two atom primitive unit cell was extracted for band structure calculations. This extraction of primitive unit cell was helpful in avoiding band folding. The effect of SOC was included in band structure calculation. Dynamical stability of YBi under the effect of hydrostatic pressure and epitaxial strain was verified with phonon dispersion calculations using Phonopy code [38]. Product of parities at TRIM points, in the presence of TRS and inversion symmetry (IS), was used to calculate $Z_2$ topological invariants. To parametrized the tight-binding (TB) Hamiltonian, we were obtained the maximally localised wannier functions (MLWF) using wannier90 code [39]. The surface band structure and WCCs were calculated with surface Green's function methods using WannierTools code [40].

## III. RESULTS AND DISCUSSIONS

### A. Structural and stability analysis, and electronic structure at ambient condition

Alike many other rare-earth monopnictides, YBi exists in a stable *rocksalt type* (*NaCl-type*) crystal structure having space group $Fm\bar{3}m$ (#225) with Y (0.5, 0.5, 0.5) and Bi (0, 0, 0) atoms as shown in Fig. 1(a). Its optimized lattice parameter is a= 6.338 Å, which is in good agreement with previous theoretical and experimental reports as mentioned in Table I.

TABLE I: Lattice parameter and Structural phase transition (SPT) of YBi.

| YBi | Previous experimental study | Previous theoretical study | Present study |
|---|---|---|---|
| Lattice parameter, $a$ (Å) | 6.2597 [41] | 6.3378 [42]; 6.345 [43]; 6.252 [44]; 6.29 [45] | 6.338 |
| SPT in GPa | --------- | 28.1 [42]; 23 [42]; 24 [44]; 23.4 [45] | 24.5 |

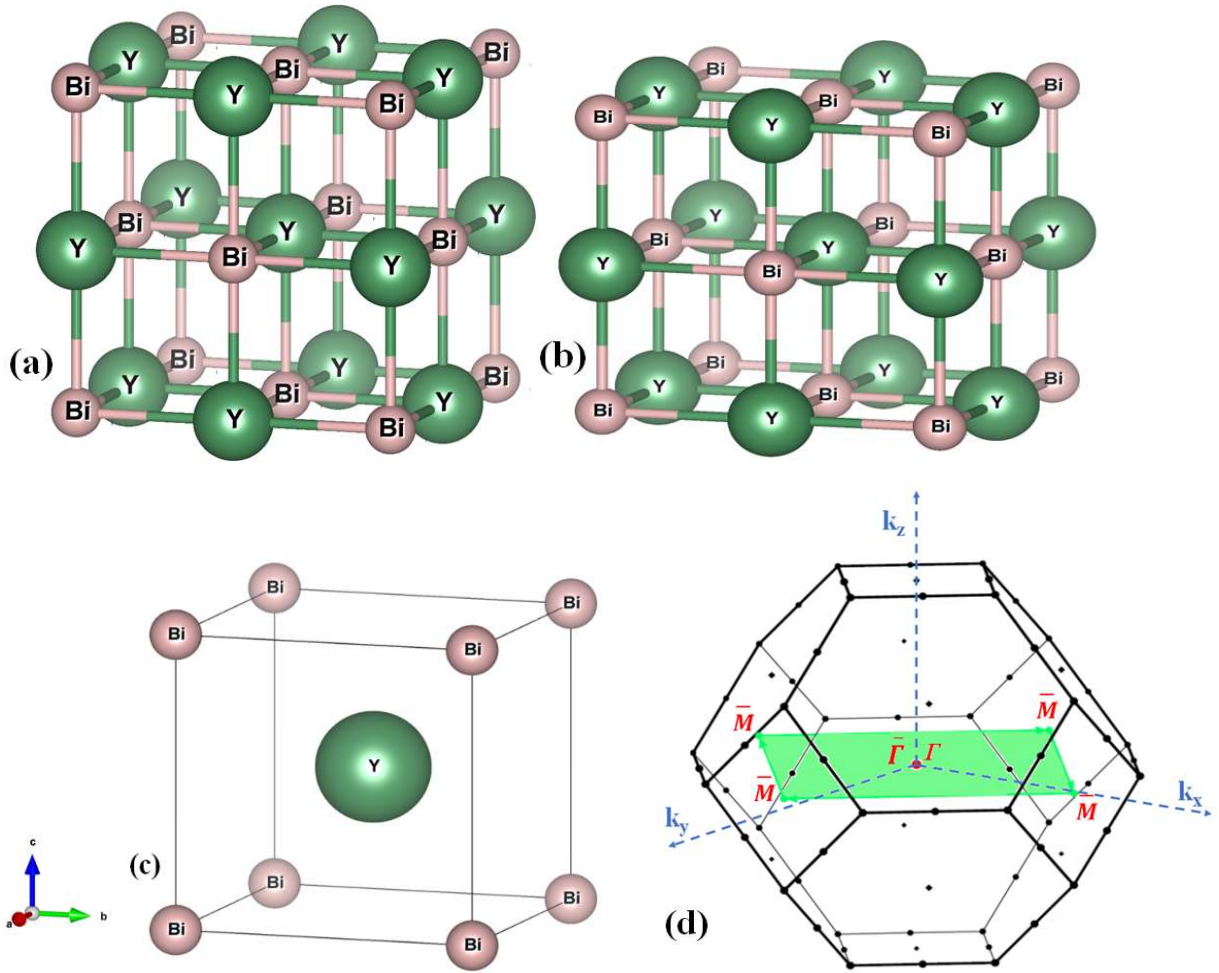

FIG. 1: Crystal structure of YBi in (a) FCC (*NaCl-type*), (b) tetragonal, (c) BCC (*CsCl-type*) structure, (d) The Brillion zone (BZ) of YBi. The shaded area (green colour) is representing the projection of the bulk Brillouin zone on the (001) surface Brillouin zone (SBZ), with symmetry points in the SBZ displayed (red colour). Here, center of BZ ($\Gamma$) and its projection in SBZ ($\bar{\Gamma}$) are coinciding.

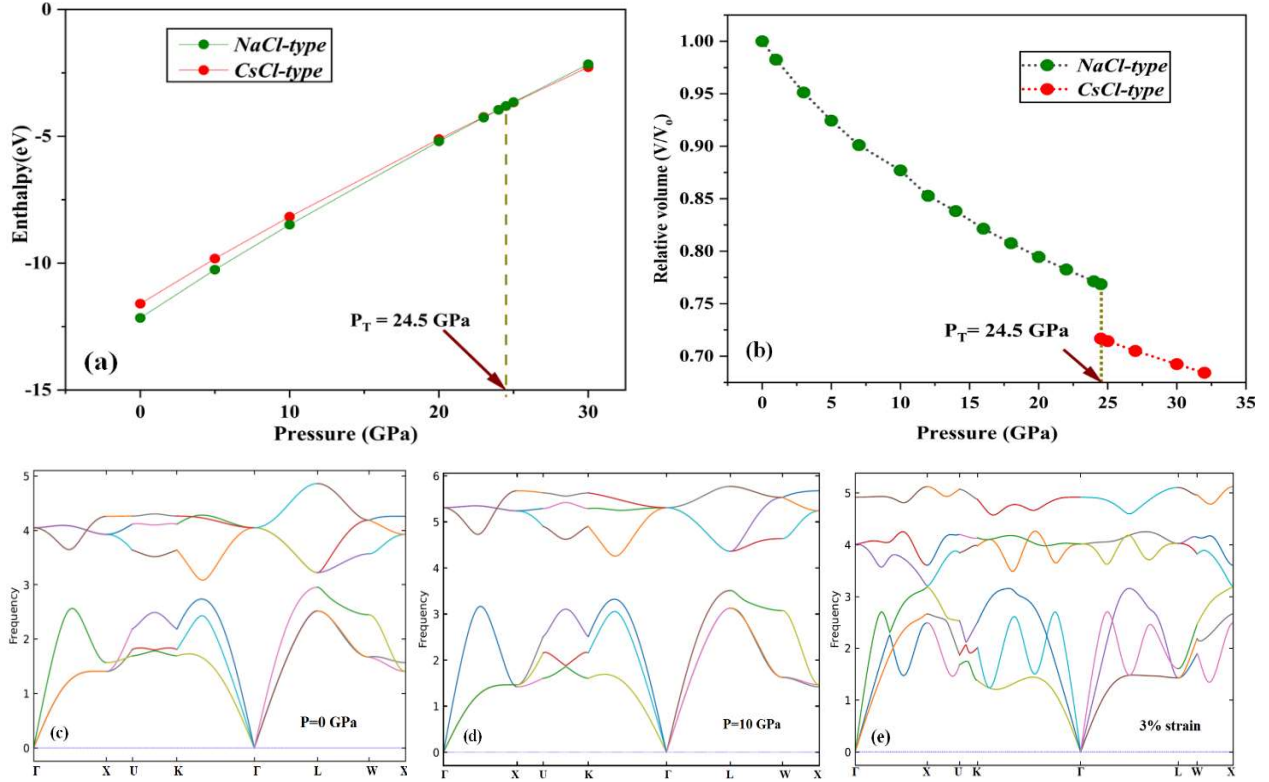

FIG. 2: (a) Enthalpy of YBi as function of pressure for *NaCl-type* to *CsCl-type* structure. (b) Variation in relative volume of YBi as a function of pressure. The phonon dispersion of YBi at (c) 0 GPa, (d) 10 GPa pressure and (e) 3% strain.

The YBi shows structural phase transition (SPT), with applied volumetric pressure, and converts to a *CsCl-type* structure (Fig. 1(c)) from the *NaCl-type* structure as shown in Fig. 2 (a). The stability of a structure at a given pressure can be accessed through Enthalpy which is defined as $H = E + PV$, where E is total energy, V is volume and P is the external pressure on unit cell. We found the SPT at around 24.5 GPa which is in good agreement with previous reports as listed in Table I. The variation in the relative volume of YBi with applied volumetric pressure is shown in Fig. 2(b). The sudden change in volume at SPT signifies a first-order phase transition resulting in change of the crystal symmetry. The *rocksalt* crystal structures have truncated octahedron Brillion zone (BZ) under equilibrium conditions, with $\Gamma$ as a center (Fig. 1(d)). The (001) plane (green colour) containing center of BZ ($\Gamma$-point) and $\bar{M}$-points at center of squares are shown in Fig.1(d). When we apply volumetric pressure, the changes in the BZ are the same in all directions and it holds its truncated octahedron shape. On the other hand, under epitaxial strain, three *X-points* along momentum axis are divided into two in-plane and one out-of-plane *Z-point*, and a distorted BZ with preserved inversion symmetry is observed. The dynamical stability of YBi under applied volumetric pressure and epitaxial strain is also analysed. As shown in Fig. 2(c-e), the phonon dispersion spectrums have no negative frequency which confirms that YBi is dynamically stable and can be realized experimentally under studied pressure and strain conditions.

To establish the true nature of YBi at ambient pressure, we have plotted the band structures with SOC using two functionals i.e., GGA-PBE and HSE06. With former, we found that YBi is topologically trivial semimetal with even (two) number of band inversions between Y-$d$ band and Bi-$p$ band at $\Gamma$-and $X$-points as shown in Fig. 3(a). A previous study has identified it as topological semimetal with a single Dirac cone at $\Gamma$-point [30]. Further, it has been reported experimentally using ARPES that YBi is topologically trivial having no Dirac cone [29]. To accurately predict the true nature of this material, we have used more accurate hybrid functional HSE06 as shown in Fig. 3(b). It can be seen the partially occupied bands at $\Gamma$-point (hole pockets) and $X$-point (electron pockets) which are acquired mostly by *6p-orbitals* of Bi and *4d-orbital* of Y, respectively. A small overlap in energy between the Y-$d$ band and Bi-$p$ band at Fermi level confirms the semimetallic nature of YBi with no band inversion in line with the experimental report [29] which can also be verified with projected density of state (PDOS) (Fig. 3(c)). This topological trivial nature of YBi is also established by absence of Dirac cone in surface states as shown in Fig. 3(d).

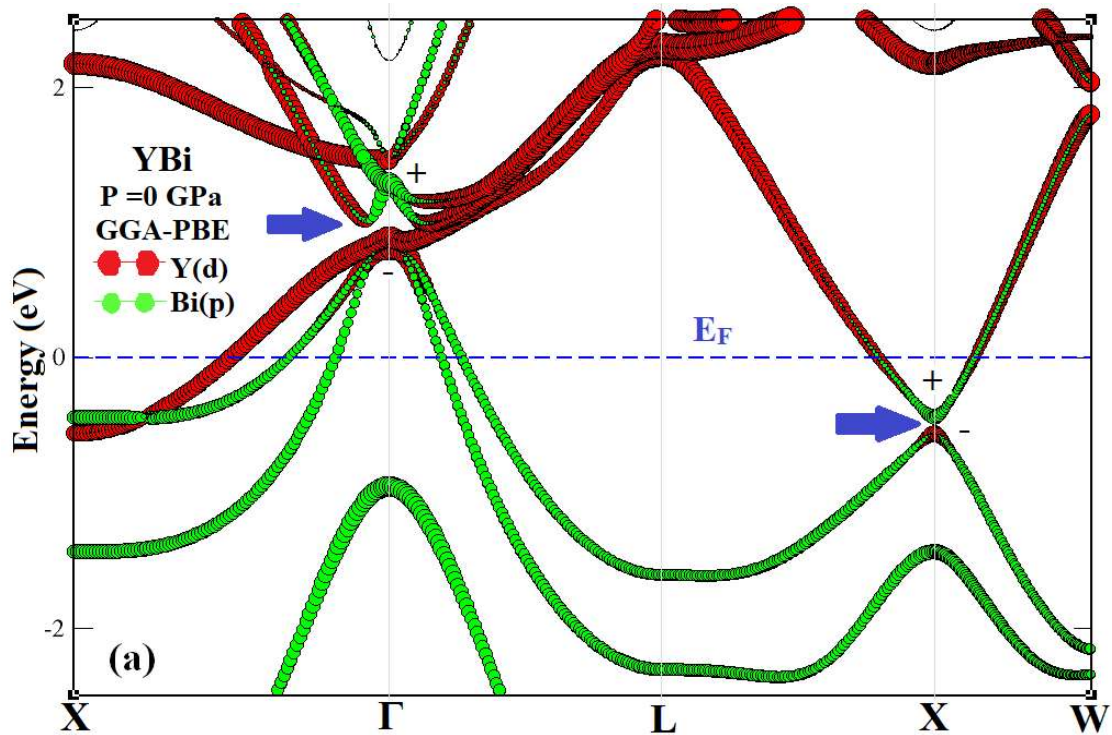

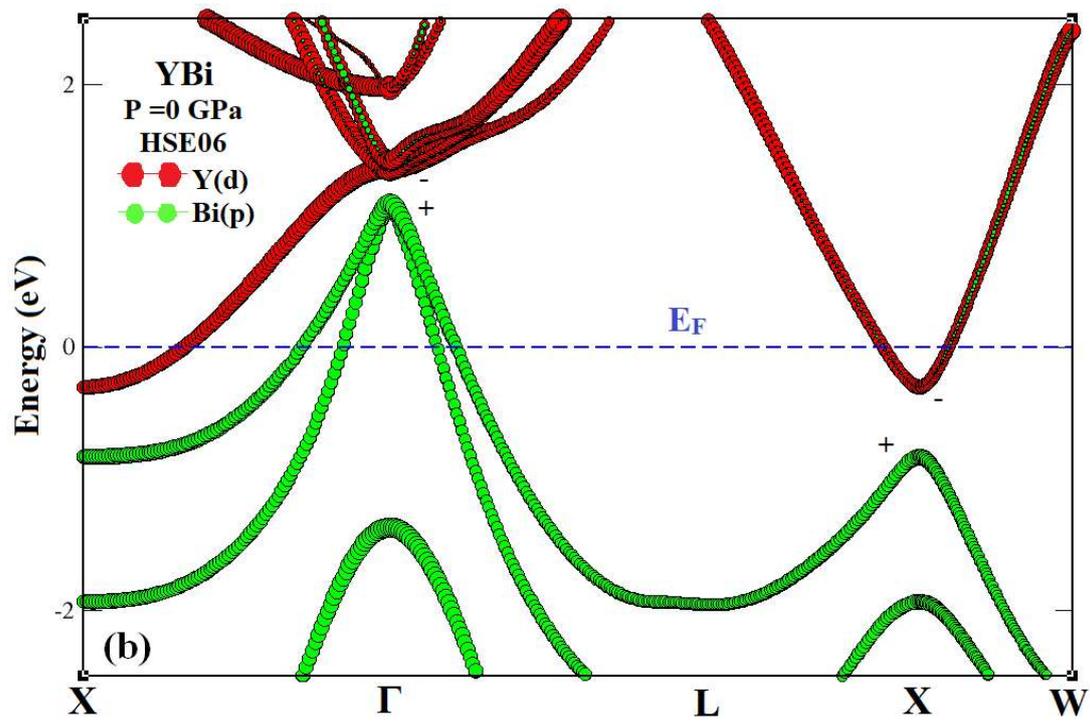

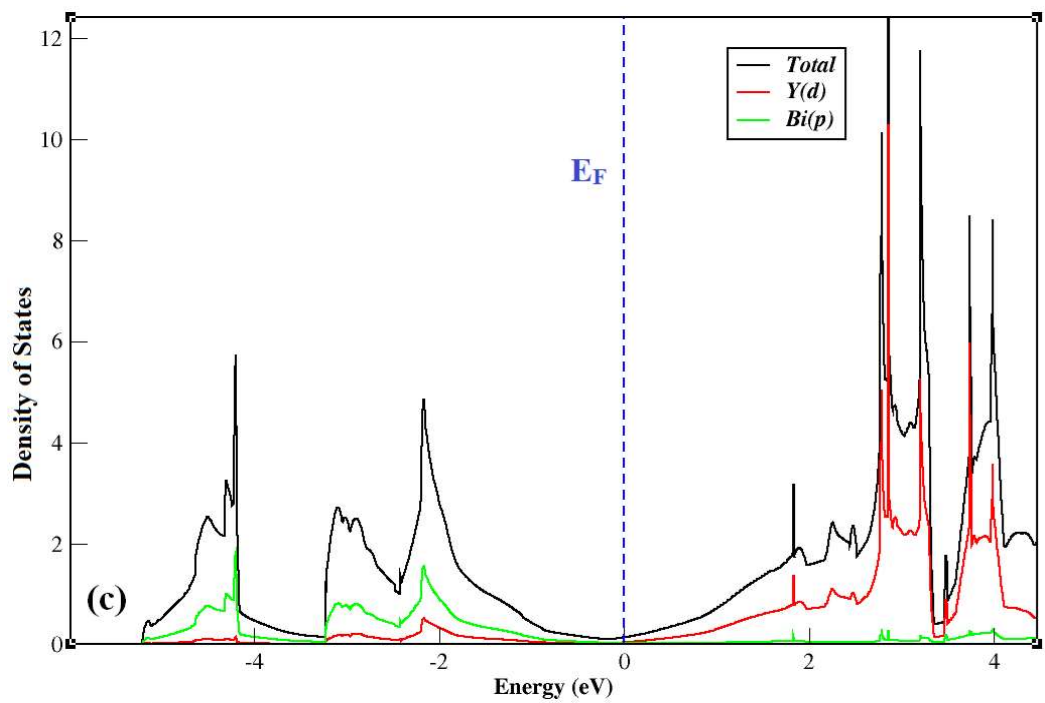

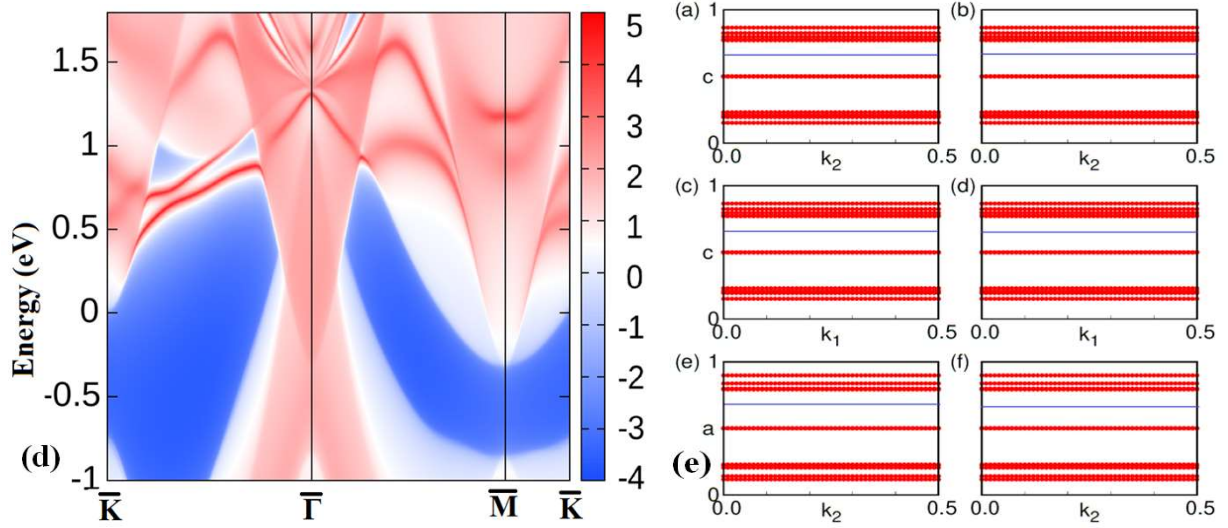

FIG. 3: The band structures of YBi with inclusion of SOC effect using (a) GGA-PBE, (b) HSE06; (c) Projected density of states; (d) The surface state and (e) Wannier charge centers (WCCs) of YBi along (001) plane.

TABLE II: The Parities of all the occupied bands at all the TRIM points in BZ of YBi at ambient pressure.

| Band No. | L | L | L | L | Γ | X | X | X | Total |
|---|---|---|---|---|---|---|---|---|---|
| 1 | - | - | - | - | - | - | - | - | + |
| 3 | - | - | - | - | - | - | - | - | + |
| 5 | - | - | - | - | - | - | - | - | + |
| 7 | - | - | - | - | + | + | + | + | + |
| 9 | + | + | + | + | - | - | - | - | + |
| 11 | + | + | + | + | - | - | - | - | + |
| 13 | + | + | + | + | - | - | - | - | + |
| Total | + | + | + | + | + | + | + | + | + |

Further, the calculation of the $Z_2$ topological invariant is performed using product of parities of bands to verify the topological nature of YBi. For three-dimensional materials, having both inversion symmetry as well as TRS, four $Z_2$ topological invariants can be calculated from the product of parities of occupied bands at TRIM points as suggested by Kane and Mele [15]. These four $Z_2$ topological invariants i.e., $v_0$; $v_1$, $v_2$, $v_3$ can be identified using relations;

$$(-1)^{\nu_0} = \prod_{m_j=0,1} \delta_{m_1 m_2 m_3} \qquad (1)$$

$$(-1)^{\nu_{i=1,2,3}} = \prod_{m_{j \neq i}=0,1 \& m_i=1} \delta_{m_1 m_2 m_3} \qquad (2)$$

where $\delta$ shows the product of parities of all occupied bands at selected TRIM points, $\nu_0$ identifies the topological phase and $\nu_1$, $\nu_2$, $\nu_3$ are used to identify the weak topological nature of YBi. Parities of the all filled energy states at ambient condition are represented in Table II. The first $Z_2$ invariant ($\nu_0$) is zero (from equation (1)) which signifies topological trivial nature of YBi.

The $Z_2$ topological invariants for a bulk material can also be obtained using Wilson loop method also [16]. In this analysis, the $Z_2$ topological invariants can be calculated with the help of evolution of WCCs [16,40,46] along six TRIM planes i.e., $k_x=0$, $\pi$; $k_y=0$, $\pi$ and $k_z=0$, $\pi$. The appearance of WCCs has been analysed for YBi using the planes that are spanned by TRIM points. Since the system exhibits TRS, one can place a random reference line across the x-axis, which corresponds to the pumping direction over half of the BZ, to figure out whether it is topologically trivial or non-trivial. The number of crossings of reference line with the evolution lines of WCCs with SOC, provide information about topological nature of YBi. If even number of crossings between reference line and WCCs takes place than it represents the trivial nature and $Z_2$ topological index have value 0. On the other hand, odd number of reference line and WCCs crossings indicates non-trivial nature with non-zero $Z_2$ topological index. The non-zero and zero values of $Z_2$ topological index in planes having $k_x$, $k_y$, $k_z = 0$ and $k_x$, $k_y$, $k_z = 0.5$, respectively, represents the strong TI with $Z_2 = (1;000)$. It can be observed in Fig. 3(e) that WCCs evolution lines have no crossing with the reference line (blue) in $k_x$, $k_y$, $k_z = 0$ and $k_x$, $k_y$, $k_z = 0.5$ planes and hence the $Z_2$ topological invariants are (0;000), which confirms the topological trivial nature of YBi at ambient pressure condition.

### B. Volumetric Pressure

After understasting the electronic structure and topological properties of YBi at ambient pressure, we now include the effect of volumetric pressure. We have examined the band structures of the YBi *rocksalt* structure using the HSE06 functional across a pressure range of 1 to 24.5 GPa. We have found no band inversion till 6.4 GPa of volumetric pressure. At 6.5 GPa, a clear band inversion has detected at *Γ- point* where *d-orbital* of Y and *p-orbital* of Bi gets inverted (Fig. 4(a)). Therefore, we can say that YBi undergoes a topological phase transition at 6.5 GPa. Further, the non-trivial topological nature of YBi can be observed with the help of (001) surface band structure as shown in Fig. 4(b). It is found that the surface states show single Dirac cone (Fig. 4(b)) at $\bar{\Gamma}$-*point* which corresponds to band inversion at *Γ-point* in bulk band structure projected SBZ. Further increase in the pressure up to 10 GPa results in another band inversion at *X-point* which can be seen in Fig. 5(a). Now, we have an even numbers of band inversions i.e., one at *Γ*- and other at *X-point*, which makes YBi a topologically trivial or weak in nature again. Bulk band structure projection on SBZ also confirms the existence of two Dirac cone at $\bar{\Gamma}$- and $\bar{M}$-*points* in surface band structure as shown in Fig. 5(b).

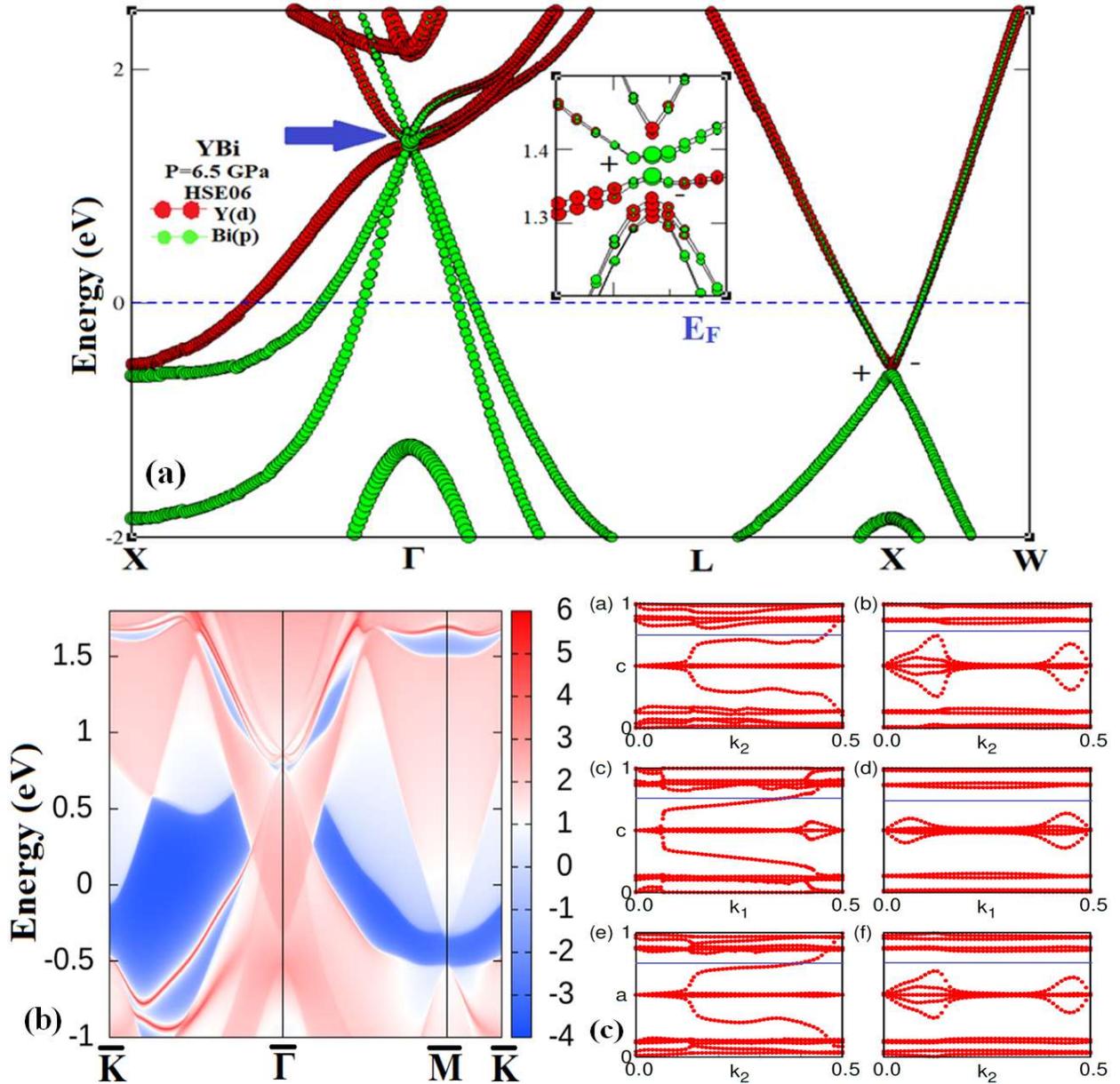

FIG. 4: (a) The band structures of YBi with inclusion of SOC effect using HSE06 functional at 6.5 GPa. (b) The surface state and (e) Wannier charge centers (WCCs) of YBi along (001) plane at 6.5 GPa.

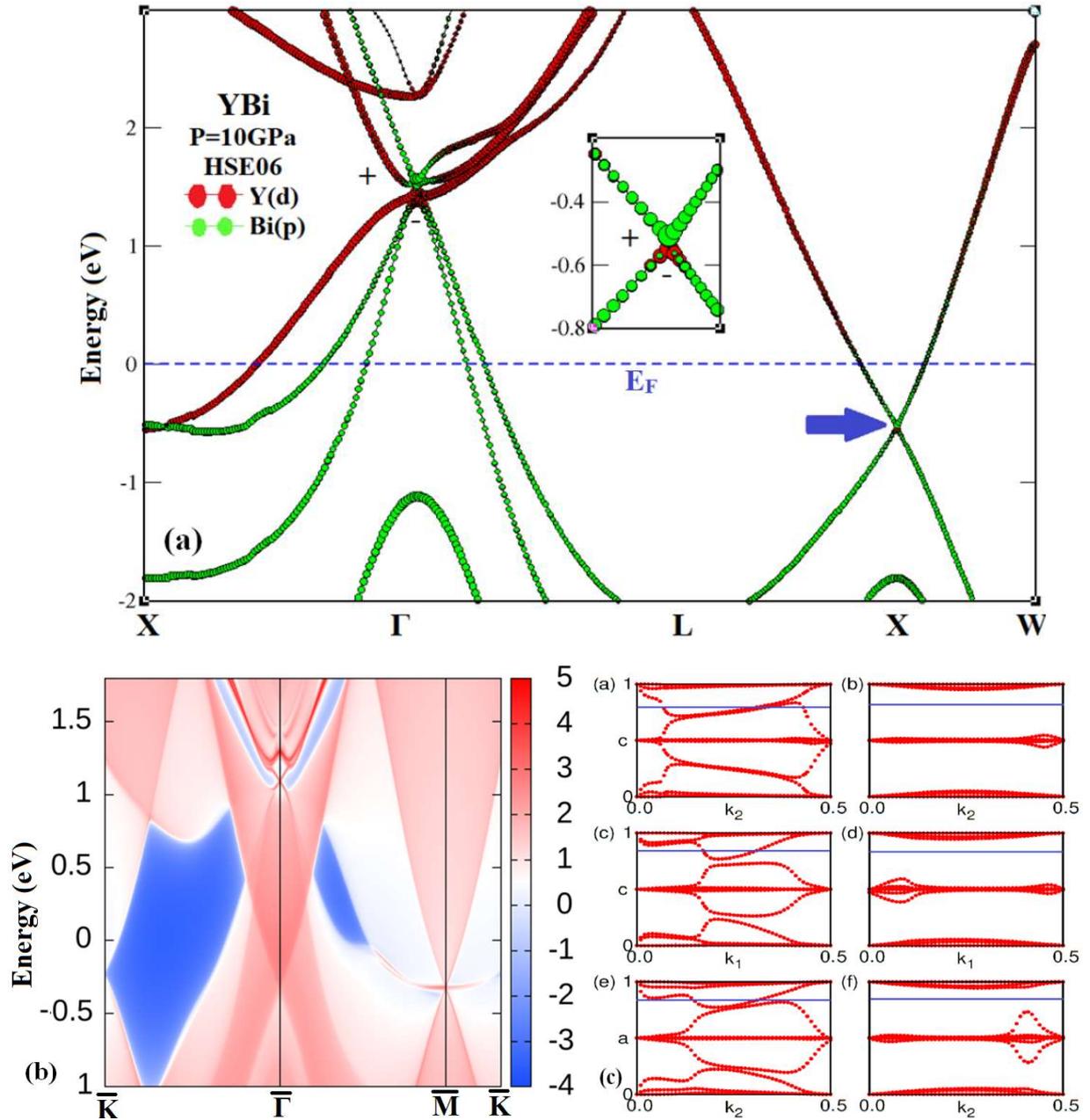

FIG. 5: (a) The band structures of YBi with inclusion of SOC effect using HSE06 functional at 10 GPa. (b) The surface state and (e) wannier charge centers (WCCs) of YBi along (001) plane at 10 GPa.

To understand the band inversion and change in parity under volumetric pressure, we have analysed the band structure evolution at $\Gamma$- and *X-points* starting from atomic energy levels and then introducing (i) octahedral field (ii) crystal field (iii) spin-orbit interaction (SOI), and (iv) pressure (Fig. 6(a-b)). We have used *pd* model for analysis of crystal field splitting in YBi [47-48]. Under the effect of applied volumetric pressure, *Y-$d_{z^2}$* orbital shifts down and *Bi-$p_{x,y}$* orbital shifts up, as expected. At critical values of the pressure 6.5 GPa and 10 GPa, respectively, band inversions take place at *Γ*- and *X-points* due to these shifts in orbitals as shown in Fig. 6(a-b). Therefore, YBi changes from a normal semimetal to a topological one. Fig. 6(c) depicts the phase diagram with respect to the different exchange-correlation functionals with SOC and pressure. The GGA-PBE and HSE06 shows band inversions at ambient pressure and 6.5 GPa, respectively.

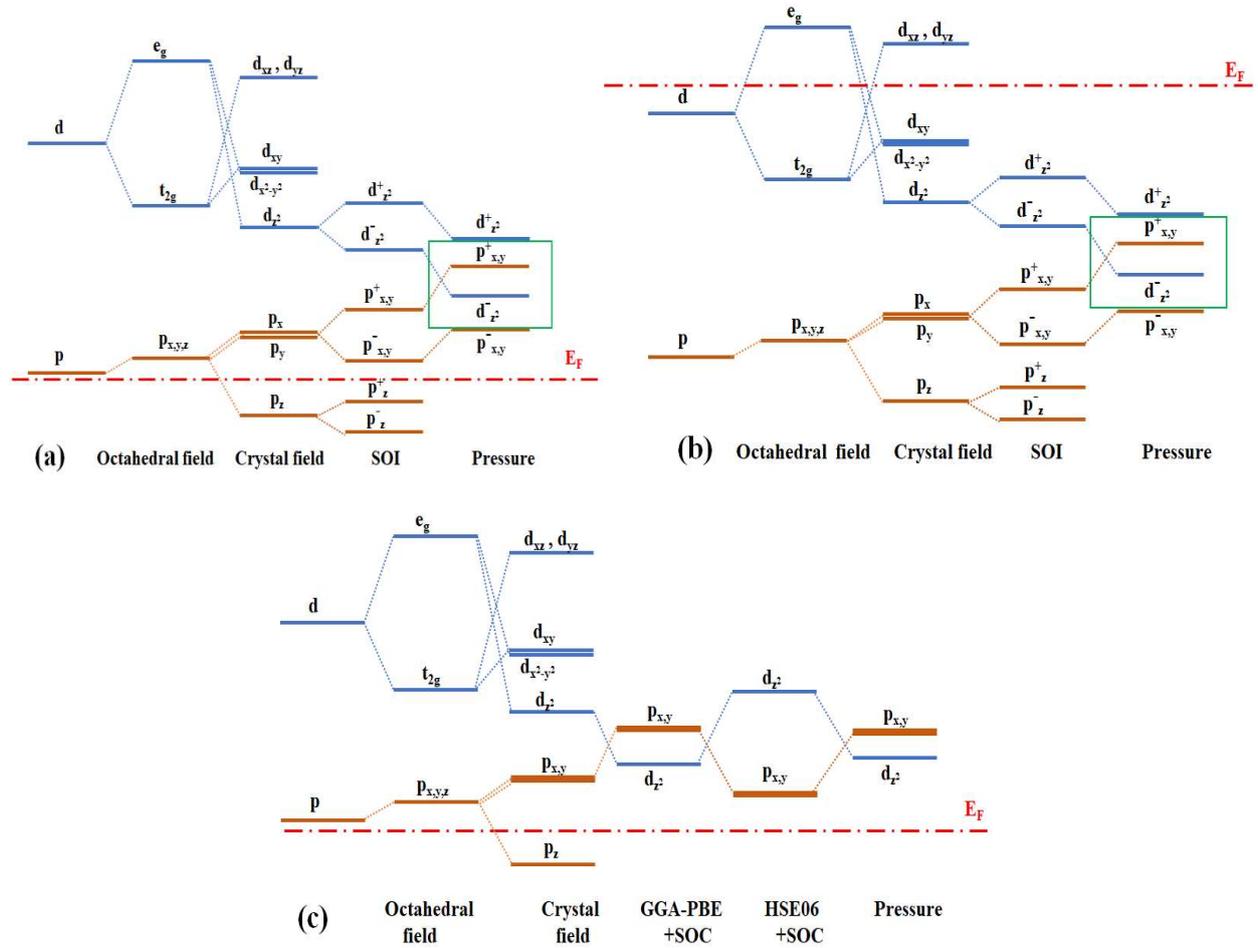

FIG. 6 The band structure evolution of YBi starting from atomic orbitals—octahedral field—crystal field splitting— SOI —applied pressure (a) at *Γ*-point (b) at *X*-point (c) with GGA-PBE and HSE06 functionals at *Γ*-point.

Parities of all the filled energy states at 6.5 GPa and 10 GPa are depicted in Tables III and IV, respectively. At first band inversion i.e., at 6.5 GPa, the parity of the highest occupied band at *Γ* is changed to positive, which switched the overall parity of YBi to negative as shown in Table III.

TABLE III: The Parities of all the occupied bands at all the TRIM points in BZ of YBi at 6.5 GPa.

| Band No. | L | L | L | L | Γ | X | X | X | Total |
|---|---|---|---|---|---|---|---|---|---|
| 1 | - | - | - | - | - | - | - | - | + |
| 3 | - | - | - | - | - | - | - | - | + |
| 5 | - | - | - | - | - | - | - | - | + |
| 7 | - | - | - | - | + | + | + | + | + |
| 9 | + | + | + | + | - | - | - | - | + |
| 11 | + | + | + | + | - | - | - | - | + |
| 13 | + | + | + | + | + | - | - | - | - |
| Total | + | + | + | + | - | + | + | + | - |

TABLE IV: The Parities of all the occupied bands at all the TRIM points in BZ of YBi at 10 GPa.

| Band No. | L | L | L | L | Γ | X | X | X | Total |
|---|---|---|---|---|---|---|---|---|---|
| 1 | - | - | - | - | - | - | - | - | + |
| 3 | - | - | - | - | - | - | - | - | + |
| 5 | - | - | - | - | - | - | - | - | + |
| 7 | - | - | - | - | + | + | + | + | + |
| 9 | + | + | + | + | - | - | - | - | + |
| 11 | + | + | + | + | - | - | - | - | + |
| 13 | + | + | + | + | + | + | + | + | + |
| Total | + | + | + | + | - | - | - | - | + |

Now, the first $Z_2$ topological invariant ($v_0$) becomes 1 using equation (1), which verifies the non-trivial nature of YBi. At second inversion (10 GPa), the parity of three *X-points* switches from positive to negative and first $Z_2$ topological invariant ($v_0$) changes 0 from 1 (equation (1)). Now, the YBi becomes either a weak topological insulator or topologically trivial insulator. To verify this, we have calculated the other three topological invariants ($v_1$, $v_2$, $v_3$) using equation (2). Table III shows that parities at three *X-points* and four *L-points* are the same, which indicates that the other three topological invariants are (0, 0, 0). So, it can be concluded that at 10 GPa, YBi shows an even number of band inversions and is topologically trivial in nature.

At 6.5 GPa volumetric pressure, the WCCs evolution lines in Fig. 4(c) cuts odd numbers of time to the reference line (blue) in $k_x$, $k_y$, $k_z$ =0 and $k_x$, $k_y$, $k_z$ =0.5 planes which confirms the topological non-trivial nature and Z$_2$ indices are (1;000). Whereas, in Fig. 5(c), even number of crossings between WCCs evolution lines and reference line (blue) can be seen for 10 GPa pressure, which again confirms the transition from non-trivial to trivial nature with Z$_2$ topological indices (0;000).

### C. Epitaxial Strain

The coherently strained films on lattice mismatched substrates can influence the electronic structure of materials by means of epitaxial strain. The implementation of molecular beam epitaxy method had successfully shown the presence of epitaxial strain induced during the growth process of rare-earth pnicitides on III-V semiconductors [49]. The III-V semiconductors [26,50] have attained compressive epitaxial strain of up to 3%, and a similar behaviour can be expected from rocksalt rare-earth monopnictides e.g., for LaSb and SnTe, respectively, 1.6% epitaxial and 1.1% out-of-plane tensile strain have been reported previously [26,27]. This induced strain may influence the charge transfer at the interface which can further affect the carrier compensation [26].

Now, in the following section, we will discuss about the topological phase transition in YBi when it is subjected to epitaxial strain. The space group symmetry of YBi is changes from $Fm\overline{3}m$ to $I4/mmm$ (Fig.1(b)) with epitaxial strain but the inversion symmetry remains preserved. Here we have demonstrated that the epitaxial strain pushes the band structure of YBi from topological trivial to non-trivial nature and thus creating an inevitable Dirac node at $\Gamma$-point.

The electronic band structure of YBi under compressive epitaxial strain is obtained along $X$-$\Gamma$-$L$-$X$-$W$ k-path as shown in Fig. 7(a). With epitaxial strain, the Y-$d$ band shifted towards the Bi-$p$ bands at $\Gamma$ and $X$ points which results in reduction of the total volume of the cell. At 3% strain, we find a band inversion at $\Gamma$-point; but still at X point, the Y-$d$ and Bi-$p$ bands continues to avoid band crossing. The band inversion at $\Gamma$-point can be seen in Fig. 7(a) and the inverted contribution of *d-orbital* of Y and *p-orbital* of Bi is shown in Fig. 7(a) (inset). To further verify the topological nontrivial nature of YBi under epitaxial strain, we computed the surface band structure along (001) plane. Since epitaxial strain causes the bulk band inversion in YBi only at the $\Gamma$-point, we found a single Dirac cone to emerge at the $\overline{\Gamma}$-point on (001) plane. The surface band structure along $M$-$\Gamma$-$M$ path of (001) plane is shown in Fig. 7(b). Unlike volumetric pressure, no Dirac cone is observed at *X-point* for epitaxial strain.

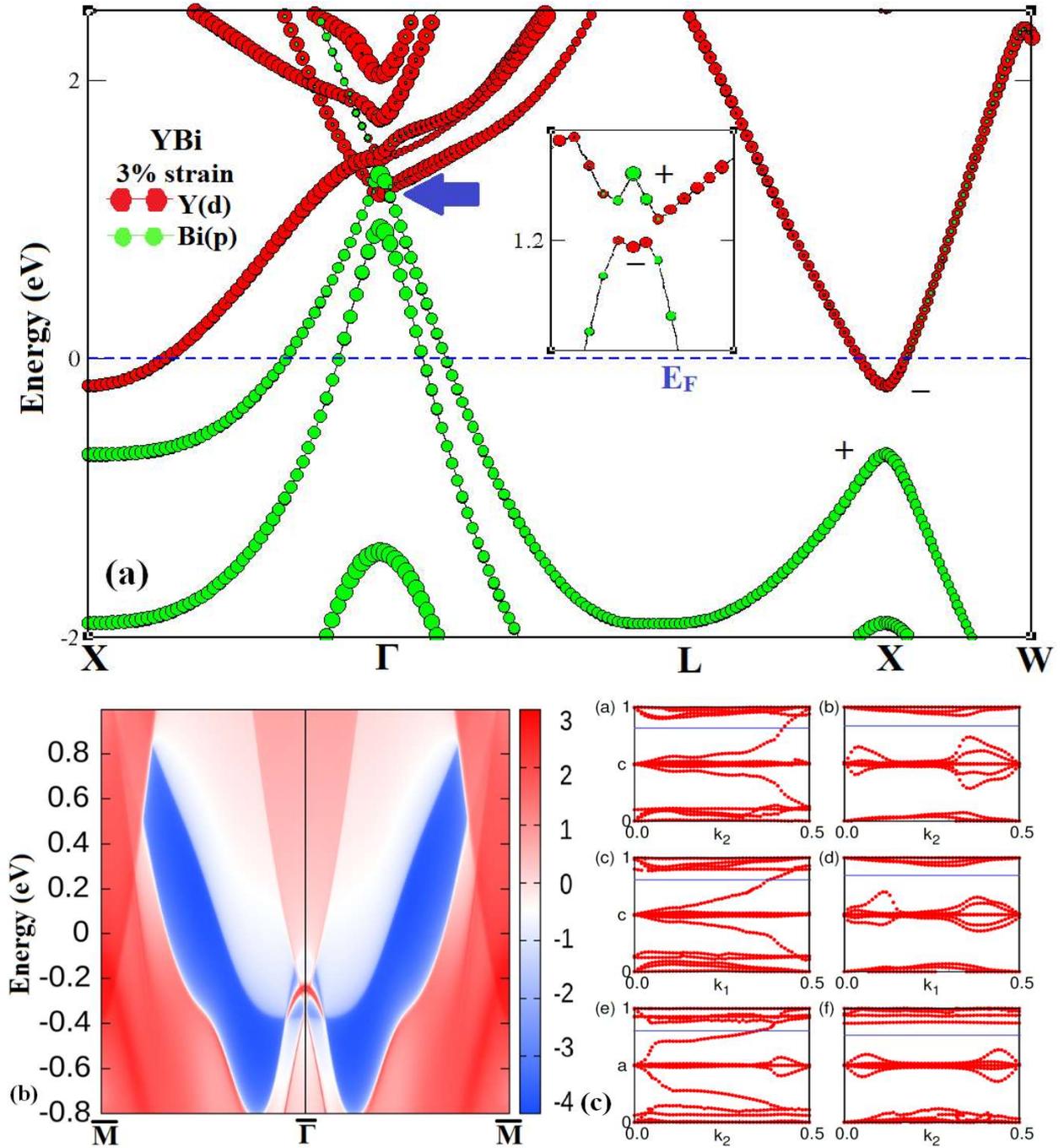

FIG. 7: (a) The band structures of YBi with inclusion of SOC effect using HSE06 functional at 3% epitaxial strain. (b) The surface state (SS) and (e) Wannier charge centers (WCCs) of YBi along (001) plane at 3% epitaxial strain.

In order to establish the occurrence of bulk band inversion under epitaxial strain and its connection with nontrivial topology in YBi, we determined the $Z_2$ topological invariant. Table V contains the parities of various occupied bands at TRIM points at an epitaxial strain of 3%. It can be observed that the parity at the $\Gamma$-point undergoes exchanged, whereas the parity at the $X$-points stays unaltered with respect to the ambient conditions (Table II). The $Z_2$ invariant change from 0 to 1 as a result of a change in parity at the *Γ-point* under epitaxial strain which is evidence of the topological non-trivial character in YBi. Moreover, the single crossing in WCCs evolution lines and reference line (blue) (Fig. 7(c)) also verifies the strong topological phase in YBi with $Z_2$= (1;000).

TABLE V: The Parities of all the occupied bands at all the TRIM points in BZ of YBi at 3% epitaxial strain.

| Band No. | L | L | L | L | Γ | X | X | X | Total |
|---|---|---|---|---|---|---|---|---|---|
| 1 | - | - | - | - | - | - | - | - | + |
| 3 | - | - | - | - | - | - | - | - | + |
| 5 | - | - | - | - | - | - | - | - | + |
| 7 | - | - | - | - | + | + | + | + | + |
| 9 | + | + | + | + | - | - | - | - | + |
| 11 | + | + | + | + | - | - | - | - | + |
| 13 | + | + | + | + | + | - | - | - | - |
| Total | + | + | + | + | - | + | + | + | - |

We have shown that YBi show topological phase transition under volumetric pressure as well as epitaxial strain. In the volumetric pressure range of 6.5 GPa to 10 GPa, YBi has non-trivial topological character and an epitaxial strain of 3% transform it from trivial to no-trivial. We have calculated the $Z_2$ topological invariant $v_0$ at different values of applied volumetric pressure as well as epitaxial strain. Fig.8(a-b) is an illustration that how the value of the first $Z_2$ topological index varies as a function of volumetric pressure and epitaxial strain.

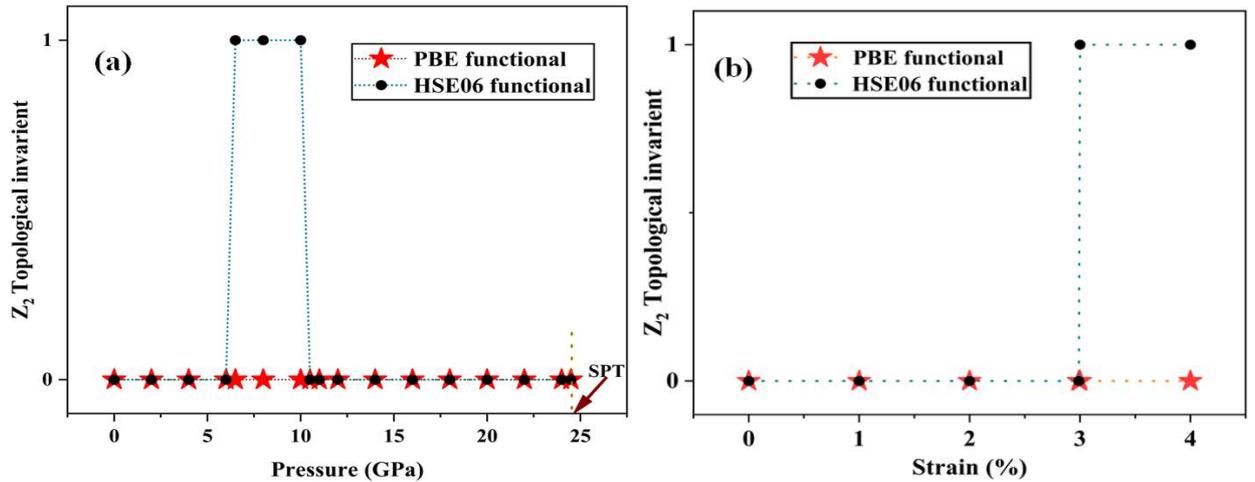

FIG. 8: The variation of First $Z_2$ topological index ($v_0$) with (a) applied volumetric pressure (b) applied epitaxial strain.

It is important to note that the rise in pressure leads to an increase in the overlap between the valence band and the conduction band of YBi, which in turn results in an increase in the carrier concentration. Since XMR counts on the carrier concentration as well as the mobility of the charge carriers, the enhancement of mobility with pressure is an effective way to analyse evolution of XMR as a function of pressure. Our results regarding topological phase transitions in YBi with 6.5 GPa of volumetric pressure can be a vital platform explore the relation between XMR and pressure. The occurrence of non-trivial phase in YBi with a relatively small epitaxial strain, which a thin film geometry can naturally has, might make it ideal candidate to probe inter-relationship between XMR and non-trivial topology.

## III. SUMMARY

We have used hybrid density functional theory to investigate the structural, electronic, and topological properties of XMR material YBi at ambient and elevated volumetric pressure and epitaxial strain. The structural and dynamical stabilities of the system have been ascertained and a structural phase transition has been predicted at 24.5 GPa. The GGA-PBE functional has overestimated the bands overlap near the Fermi level and an even number of band inversions have been observed. The hybrid functional HSE06 has accurately predicted the topologically trivial semimetallic nature of YBi which agrees with existing experimental report. The YBi has undergone a topological phase transition at 6.5 GPa of volumetric pressure and 3% of epitaxial strain. The non-zero values of $Z_2$ topological index, calculated with the help of product of parities of all the occupied bands at TRIM points and evolution of WCCs, have confirmed these topological phase transitions. The *d-orbital* of Y and *p-orbital* of Bi have mainly contributed near the Fermi level and take part in topological band inversion of YBi. The existence of single Dirac cone on plane (001) has confirmed the non-trivial nature of YBi. A rise in volumetric pressure (10 GPa) has make it trivial again which has been verified with even number of Dirac cone on (001) plane. The $Z_2$ topological index has switched from 1 to 0 at 10 GPa, which has also observed in evolution of WCCs. A small epitaxial strain of 3%, which can arise due to lattice-substrate mismatch during coherent growth of thin films of YBi, can be an opportunity to interrelate non-trivial topology, electron-hole compensation and XMR in rare-earth monopnictides.

## ACKNOWLEDGEMENTS

One of the authors (Ramesh Kumar) would like to thank Council of Scientific and Industrial Research (CSIR), Delhi, for financial support. All the authors acknowledge the National

Supercomputing Mission (NSM) for providing computing resources of 'PARAM SEVA′ at IIT, Hyderabad, which is implemented by C-DAC and supported by the Ministry of Electronics and Information Technology (MeitY) and Department of Science and Technology (DST), Government of India.